\documentclass[cits]{PoS}
\usepackage{cite}
\usepackage{amsmath}
\usepackage[english]{babel}
\def\a{\alpha}

\def\d{\delta}

\def\l{\lambda}
\def\m{\mu}

\def\t{\tau}

\def\y{\eta}

\def\f{\varphi}

\def\mtil{\widetilde{m}}

\newcommand{\GeV}{\,{\rm GeV}}

\newcommand{\beq}{\begin{equation}}
\newcommand{\eeq}{\end{equation}}
\newcommand{\bea}{\begin{eqnarray}}
\newcommand{\eea}{\end{eqnarray}}
\newcommand{\ba}{\begin{array}}
\newcommand{\ea}{\end{array}}
\newcommand{\eps}{\epsilon}

\newcommand{\gsim}{\lower.7ex\hbox{$\;\stackrel{\textstyle>}{\sim}\;$}}
\newcommand{\lsim}{\lower.7ex\hbox{$\;\stackrel{\textstyle<}{\sim}\;$}}
\newcommand{\baz}{\begin{array}{cc}}
\newcommand{\bad}{\begin{array}{ccc}}

\newcommand{\dma}{\mbox{$\Delta m^2_{\rm A}$}}

\title{On the Interplay Between the ``Low'' and ``High'' Energy CP-Violation in Leptogenesis}

\ShortTitle{``Low'' and ``High'' Energy
CP-Violation in Leptogenesis}

\author{\speaker{Emiliano MOLINARO}\\
        SISSA and INFN-Sezione di Trieste, Trieste, Italy\\
        E-mail: \email{molinaro@sissa.it}}

\author{Serguey T. PETCOV\thanks{Also at: 
Institute of Nuclear Research and Nuclear Energy, Bulgarian Academy of Sciences 1784 Sofia, Bulgaria.}\\
        SISSA and INFN-Sezione di Trieste, Trieste, Italy\\
        IPMU, University of Tokyo, Tokyo, Japan\\
        E-mail: \email{petcov@sissa.it}}

\abstract{The CP-violation necessary for the generation of the
baryon asymmetry of the Universe in the ``flavoured''
leptogenesis scenario can arise from the ``low energy'' PMNS
neutrino mixing matrix and/or from the ``high energy'' part of
neutrino Yukawa couplings, which can mediate CP-violating
phenomena only at some high energy scale. The possible interplay
between these two types of CP-violation is discussed. The type I
see-saw model with three heavy right-handed Majorana neutrinos
having hierarchical spectrum is considered. The analysis shows that there
exist regions in the leptogenesis parameter space
where the relevant ``high energy'' phases have large CP-violating
values, but the purely ``high energy'' contribution to the baryon asymmetry plays
a subdominant/suppressed role in the production of baryon asymmetry
compatible with the observations and one can have successful leptogenesis if the
requisite CP-violation is provided by the Majorana phase(s) in the
neutrino mixing matrix.}

\FullConference{European Physical Society Europhysics Conference on High Energy Physics\\
         July 16-22, 2009\\
         Krakow, Poland}

\begin{document}

 We discuss thermal ``flavoured'' \cite{davidsonetal}
leptogenesis \cite{FY} in the general framework in which
the CP-violation required to reproduce the experimental value \cite{Dunkley:2008ie} of
the matter-antimatter asymmetry of the Universe, $Y_B$, is
provided both by the ``low energy'' Majorana and/or Dirac
CP-violating phases in the neutrino mixing matrix and by the ``high
energy'' phases which can be present in the matrix of neutrino
Yukawa coupling, $\l$, and can mediate CP-violating processes only
at some ``high'' energy scale. The scheme which we consider is the
non-supersymmetric type I see-saw model \cite{seesaw} with three
heavy right-handed (RH) Majorana neutrinos, $N_j$, having masses
$M_j$ with hierarchical spectrum, $M_1\ll M_{2} \ll M_{3} $. 
In the thermal leptogenesis
scenario, the CP-violating asymmetry relevant for leptogenesis in
the case of hierarchical heavy Majorana neutrino masses, is
generated in out-of-equilibrium decays of the lightest RH
neutrino, $N_1$. The latter is produced by thermal scattering
after inflation. The results presented are based on
\cite{Molinaro:2008rg,Molinaro:2008cw}.

  In the basis in which the Majorana mass matrix of the
RH neutrinos and the matrix of the charged lepton Yukawa
couplings are diagonal, the only source of
CP-violation in the lepton sector is the matrix of neutrino Yukawa
couplings $\l$. The orthogonal parametrization of $\l$
\cite{Casas01}, involving a complex orthogonal matrix $R$, allows
to relate in a simple way $\l$ with the neutrino mixing matrix
$U$: $\lambda = (1/v)\sqrt{M} \, R\, \sqrt{m} \, U^{\dagger}$,
where $M$ and $m$ are diagonal matrices formed by the masses $M_j
> 0$ and $m_k \geq 0$ of $N_j$ and of the light Majorana neutrinos
$\nu_k$ respectively, $j,k=1,2,3$, and $v = 174$ GeV is the vacuum
expectation value of the Higgs doublet field. This parametrization
permits to  investigate the combined effect of the CP-violation
due to the ``low energy'' neutrino mixing matrix $U$ and the
CP-violation due to the ``high energy'' matrix $R$ in the
generation of the baryon asymmetry in ``flavoured'' leptogenesis.
The PMNS matrix $U$ is present in the weak charged lepton current
and can be a source of CP-violation in, e.g. neutrino oscillations
at ``low'' energies $E \sim M_Z$. The matrix $R$ does not affect the ``low'' energy neutrino mixing
phenomenology. The two matrices $U$ and $R$ are, in general,
independent. It should be noted, however, that in certain specific
cases (of, e.g. symmetries and/or texture zeros) of the matrix
$\lambda$ of neutrino Yukawa couplings, there can exist a relation
between (some of) the CP-violating phases in $U$ and (some of) the other
CP-violating parameters in $\l$ (see, e.g.
\cite{PRST05,PSMoriond06,Hagedorn:2009jy}). 

  We have analyzed certain aspects of the possible interplay between the ``low energy''
CP-violation due to the Dirac and/or Majorana CP-violating phases
in the PMNS matrix $U$, and the ``high energy'' CP-violation
originating from the matrix $R$, in ``flavoured'' leptogenesis. We have
looked for regions of the leptogenesis parameter space where the
relevant (high energy) phases of $R$ ($R-$phases) have large CP-violating
values, but the purely ``high energy'' contribution in $Y_B$ plays
a subdominant role in the production of baryon asymmetry
compatible with the observations. The requisite dominant term in
$Y_B$ can arise due to the ``low energy'' CP-violation in the
neutrino mixing matrix $U$.

   We work in the ``two flavour'' regime
~\cite{davidsonetal}, in which the $\t$ flavour
interactions are in thermal equilibrium and the Boltzmann
evolution of the CP-asymmetry in the $\tau$ lepton charge,
$\eps_\t$, is distinguishable from the evolution of the
($e+\m$)-flavour asymmetry $\eps_2\equiv\eps_e+\eps_\m$. This
regime is realized at temperatures $10^9\,\GeV\lsim T\sim M_1\lsim
10^{12}\,\GeV$. In this study we neglect the effects of the
lightest neutrino mass in thermal leptogenesis \cite{MPST07} and we work for simplicity in a regime in
which the heaviest RH neutrino is almost decoupled. Both the normal hierarchical (NH) \cite{Molinaro:2008rg}
and inverted hierarchical (IH) \cite{Molinaro:2008rg,Molinaro:2008cw} light neutrino 
mass spectrum is analyzed in detail. In
the following we concentrate on the case in which the light
neutrino mass spectrum has inverted hierarchy.
\begin{figure}[t!!]
\begin{center}
\begin{tabular}{cc}
\includegraphics[width=7cm,height=6.5cm]{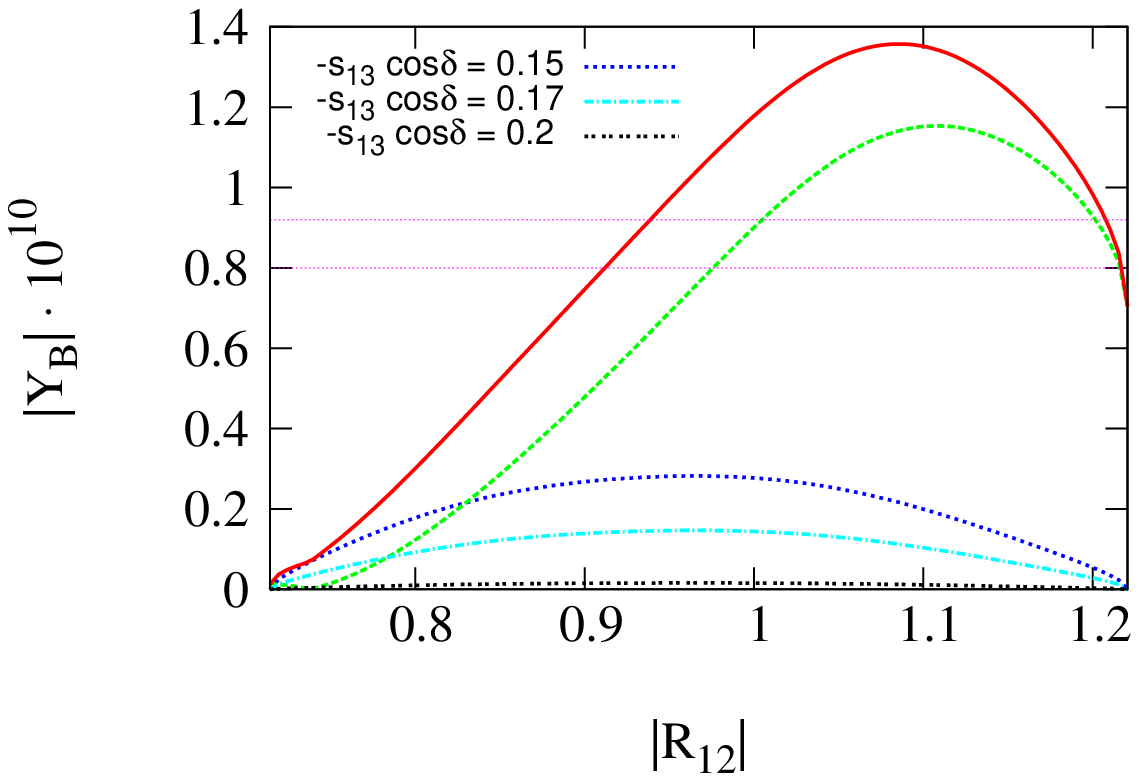} &
\includegraphics[width=7cm,height=6.5cm]{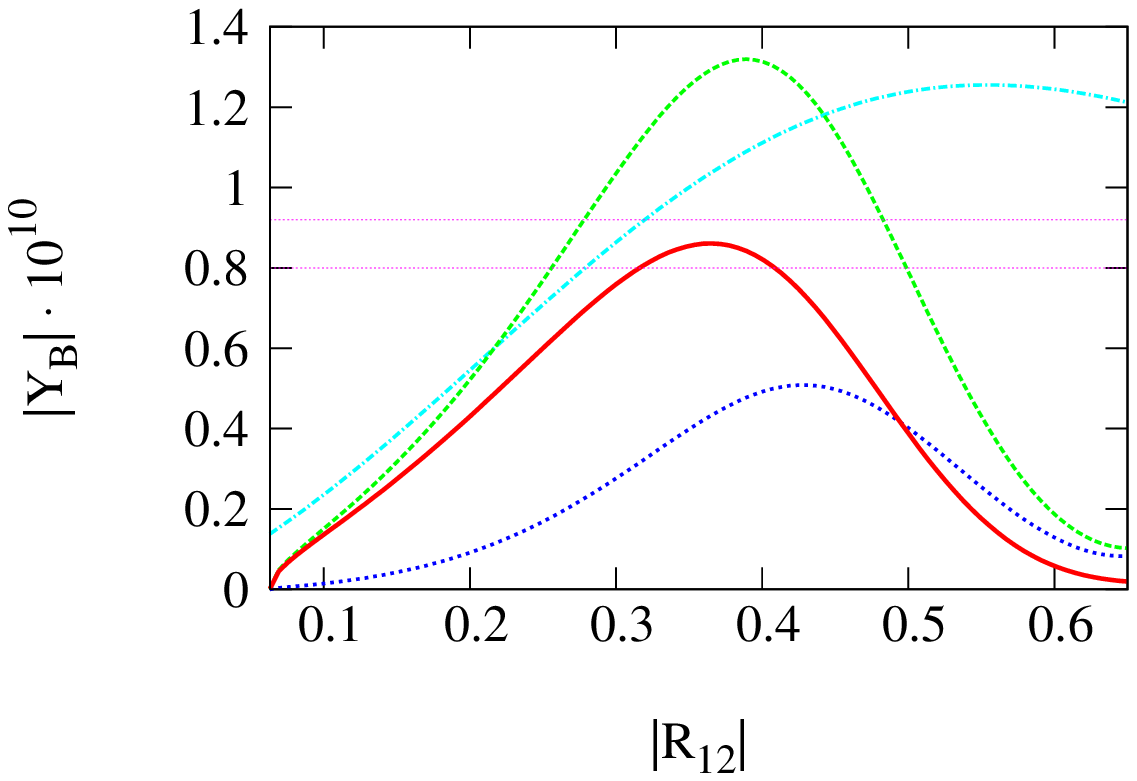}
\end{tabular}
 \caption{
\label{IH_YB_a21_0.5a} The dependence of the ``high energy'' term
$|Y^0_B A_{\rm HE}|$ (blue line), the ``mixed'' term $|Y^0_B
A_{\rm MIX}|$ (green line) and of the total baryon asymmetry
$|Y_B|$ (red line) on $|R_{12}|$ in the case of IH spectrum,
CP-violation due to the Majorana phase
$\a_{21}$ 
and $R$-phases, for $R_{13}=0$, $\alpha_{21} = \pi/2$, $M_1 = 10^{11}$ GeV and
  $i)$ $|R_{11}| = 0.7$ and $(-s_{13}\,\cos\d)=0.15,~0.17,~0.20$ (left panel); 
$ii)$ $|R_{11}| \cong 1$ and $s_{13}=0$ 
(right panel). The light-blue curve in the right panel represents the
dependence of $Y_B$ on $|R_{12}|$ for the given PMNS parameters
and CP-conserving matrix R, with 
$R_{11} R_{12}\equiv ik|R_{11}R_{12}|$, $k=-1$ and
$|R_{11}|^2-|R_{12}|^2=1$. The
horizontal lines indicate the allowed range of $|Y_B|$,
$|Y_B|=[8.0,9.2]\times 10^{-11}$.
}
\end{center}
\end{figure}
%
In this regime, the baryon asymmetry $Y_B$ can be written \cite{Molinaro:2008cw} 
as a function of $\eps_\t$ only, like in the case of the matrix
$R$ satisfying the CP-invariance constraints \cite{PPRio106}:
\begin{eqnarray}
Y_B & = &-\frac{12}{37}\frac{\eps_\t}{g_*} \left(\,
\y\left(\frac{390}{589}\mtil_{\t}\right)\,
- \,\y\left(\frac{417}{589}\mtil_2\right)\right)\nonumber\\
    &\equiv& Y^0_B(A_{\rm HE}+A_{\rm MIX})\label{YB_2}
\end{eqnarray}
\noindent where $A_{\rm HE(MIX)}\equiv C_{\rm
HE(MIX)}(\y(0.66\mtil_{\t})-\y(0.71\mtil_2))$,
$\y(0.66\mtil_{\t})$ and $\y(0.71\mtil_2))$ being the efficiency
factors for
\noindent the asymmetries $\eps_\t$ and $\eps_2$ (see
\cite{davidsonetal}), and
\begin{eqnarray}
Y^0_B & \cong &
3\times 10^{-10}\,\left ( \frac{M_1}{10^9~{\rm GeV}}\right )\,
\left (\frac{\sqrt{\dma}}{5\times 10^{-2}~{\rm eV}}\right )\,,
\label{Y0B}
\end{eqnarray}
\begin{eqnarray}
C_{\rm HE}\, = \, G_{11}\, \sin 2\tilde{\f}_{11} \left
[\,|U_{\t1}|^2-|U_{\t2}|^2\,\right ]\,, \label{CHE1}
\end{eqnarray}
\begin{eqnarray}
C_{\rm MIX}\, \cong \, 2\, G_{12}\,
\sin(\tilde{\f}_{11}+\tilde{\f}_{12})\,{\rm
Re}(U_{\t1}^*U_{\t2})\,, \label{CHM}
\end{eqnarray}
\noindent where $G_{11}\equiv|R_{11}|^2/(|R_{11}|^2+|R_{12}|^2)$,
$G_{12}\equiv|R_{11}R_{12}|/(|R_{11}|^2+|R_{12}|^2)$ . In Eq.
(\ref{YB_2}), $Y^0_B A_{\rm HE}$ is the ``high energy'' term which
vanishes in the case of a CP-conserving matrix $R$ ($R_{1j}$ real or purely imaginary \cite{PPRio106}), 
while $Y^0_B A_{\rm MIX}$ is a ``mixed'' term  which, in contrast to
$Y^0_BA_{\rm HE}$, does not vanish  when $R$ conserves CP: it
includes the ``low energy'' CP-violation, $e.g.$ due to the Majorana
phase $\alpha_{21}$ in the neutrino mixing matrix. 
In order to have
CP-violation due to the  Majorana phase $\a_{21}$, both  ${\rm
Im}(U_{\t1}^*U_{\t2})$ and ${\rm Re}(U_{\t1}^*U_{\t2})$ should be
different from zero.

 Using the formalism described above,
we have analyzed in detail the interplay between the CP-violation arising
from the ``high energy'' $R-$phases  and
the ``low energy'' CP-violating Dirac and/or Majorana phases in
the neutrino mixing matrix, as well as the relative contributions
of the ``high energy'' and the ``mixed'' terms $Y^0_B A_{\rm HE}$
and $Y^0_B A_{\rm MIX}$ in $Y_B$. We have found that there exist
large regions of the corresponding leptogenesis parameter space
where the ``high energy'' contribution to $Y_B$ is subdominant or
even strongly suppressed. Some results are illustrated in Fig.1.
Strong suppression 
of $Y_{HE}$ can arise thorough the factor $(|U_{\t1}|^2-|U_{\t2}|^2)$
(independently of the values of
the other leptogenesis parameters)
provided $(-\sin\theta_{13}\cos\d) \gsim 0.15$ (Fig.1 left panel). 
This result is valid in all the parameter 
space compatible with successful leptogenesis, where
 the ``high energy'' CP-violating phases are not
necessarily small, but reproducing the observed value of the
baryon asymmetry is problematic (or can even be impossible)
without a contribution due to the CP-violating phases in the PMNS
matrix. The ``high energy''contribution can be subdominant also in
the case of $\sin\theta_{13}= 0$ (Fig.1 right panel). This possibility can be realised
for values of the Majorana phase in the PMNS matrix $0 <
\alpha_{21} \lsim 2\pi/3$  and roughly in half of the parameter
space spanned by the relevant elements of the $R$ matrix. In both
cases the observed value of the baryon asymmetry can be reproduced
for values of the lightest RH Majorana neutrino mass lying in the
interval $5\times 10^{10}~{\rm GeV} \lsim M_1 \lsim 7\times
10^{11}$ GeV. 

  The results obtained in \cite{Molinaro:2008rg,Molinaro:2008cw} 
show that in the ``flavoured'' leptogenesis scenario, the
contribution to $Y_B$ due to the ``low energy'' CP-violating
Majorana and Dirac phases in the neutrino mixing matrix, in
certain physically interesting cases can be indispensable for the generation of the observed
baryon asymmetry of the Universe even in the presence of ``high
energy'' CP-violation, generated by additional physical phases in
the matrix of neutrino Yukawa couplings.

\end{document}